\documentclass[aps,prb,amsmath,amssymb,groupedaddress,twocolumn,superscriptaddress,showpacs]{revtex4}

\usepackage{graphicx}
\usepackage{bm}
\usepackage[dvips]{color}

\renewcommand{\vec}[1]{\mathbf{#1}}
\renewcommand{\Re}{{\operatorname{Re}}}

\begin{document}

\title{Models for the magnetic ac susceptibility of
granular superferromagnetic CoFe/Al$_2$O$_3$}

\author{O.~Petracic}
\email[e-mail:]{opetr@kleemann.uni-duisburg.de}%
\affiliation{Physics Department, University of California, San
Diego, La Jolla, California 92093} \affiliation{Angewandte Physik,
Universit\"at Duisburg-Essen, 47048 Duisburg, Germany}
\author{A.~Glatz}
\email[e-mail:]{ang@thp.uni-koeln.de}%
\affiliation{Materials Science Division, Argonne National
Laboratory, Argonne, Illinois 60439} \affiliation{Institut f\"ur
Theoretische Physik, Universit\"at zu K\"oln, 50937 K\"oln,
Germany}
\author{W.~Kleemann}
\affiliation{Angewandte Physik, Universit\"at Duisburg-Essen,
47048 Duisburg, Germany}

\date{\today}

\begin{abstract}
The magnetization and magnetic ac susceptibility, $\chi = \chi' -
i \chi''$, of superferromagnetic systems are studied by numerical
simulations. The Cole-Cole plot, $\chi''$ vs. $\chi'$, is used as
a tool for classifying magnetic systems by their dynamical
behavior. The simulations of the magnetization hysteresis and the
ac susceptibility are performed with two approaches for a driven
domain wall in random media. The studies are motivated by recent
experimental results on the interacting nanoparticle system
Co$_{80}$Fe$_{20}$/Al$_{2}$O$_{3}$ showing superferromagnetic
behavior. Its Cole-Cole plot indicates domain wall motion dynamics
similarly to a disordered ferromagnet, including pinning and
sliding motion. With our models we can successfully reproduce the
features found in the experimental Cole-Cole plots.
\end{abstract}

\pacs{75.60.-d, 75.75.+a, 75.40.Gb, 75.40.Mg}

\maketitle


\section{\label{sec-intro}Introduction}
The physics of interacting ferromagnetic (FM) nanoparticles is a
vivid topic of modern magnetism research. This also applies to the
study of the reversal dynamics in thin FM films. The first
subject, the properties of interacting FM nanoparticles, is
investigated by many groups focusing either on the preparation
(e.g., Ref.~\onlinecite{dorm-acp-97, krish-s-01}) or the magnetic
properties (e.g. Refs.~\onlinecite{dorm-acp-97, batt-jpd-02,
djur-prl-97, wood-prl-01, luis-prl-02, sahoo-apl-03}). Numerous
theoretical studies were perfomed in order to understand the
observed phenomena or to explore possible new effects (e.g.
Refs.~\onlinecite{ander-prb-97, ulri-prb-03, jensen-njp-03,
porto-epj-02, kech-prb-98}).

While individual single-domain FM nanoparticles exhibit
superparamagnetic (SPM) behavior \cite{neel, brown, dorm-acp-97,
garc-acp-00}, interacting ensembles lead to very different kinds
of phenomena depending on the type and strength of interactions.
Dipolar interactions become relevant since the magnetic moment
e.g. for particles with diameter 5nm is of the order $5000 \mu_B$,
while the inter-particle distances are of the order $1-10$~nm. The
simple formula for the mean dipolar energy of a particle to one
neighbor, $E_{d-d}/k_B=(\mu_0/4 \pi k_B) \mu^2/D^3$, yields
already $16$~K for $D=10$~nm. Considering many neighbors and
shorter distances it is obvious, that the effects of dipolar
interaction can be observed even at temperatures of the order
$100$~K. In addition, several other types of interactions are
proposed, i.e., higher order multipole terms of dipolar
\cite{poli-prb-02, russ-jap-01}, tunneling exchange
\cite{kond-prl-98} or even retarded van der Waals interactions
\cite{cham-prb-02}. Independent from the still open question which
interactions are relevant, one can summarize, that essentially
three different kinds of phenomena occur\cite{dorm-jmmm-98}:

For large inter-particle distances, and hence a small
concentration of particles, the (dipolar) interaction is only a
perturbation to the individual particle behavior and no collective
behavior is found.\cite{dorm-jmmm-98, luis-prl-02,
joensson-prb-01} For intermediate concentrations a superspin glass
(SSG) phase is encountered. In this case the particle moments
(superspins) collectively freeze into a spin glass phase below a
critical temperature, $T_g$.\cite{dorm-jmmm-98, djur-prl-97,
mamiy-prl-99, muro-prb-99, sahoo-prb-03} For even higher
concentrations a superferromagnetic (SFM) state is found. It is
characterized by a ferromagnetic arrangement of the
moments.\cite{moerup-jmmm-83, hausch-prb-98, schein-prl-96,
klee-prb-01, puntes-natm-04, baku-jmmm-04} The magnetic dynamic
behavior resembles at the first glance that of the SSG case, but
actually shows features of domain wall motion similar to an impure
ferromagnet as will be discussed below.\cite{chen-prl-02} Here one
should mention that also additional types of collective ordering
are proposed in literature, e.g., the correlated superspin glass
state (CSSG),\cite{binns-prb-02,loeff-prl-00} and also that the
effects of surface spin disorder may become
significant.\cite{koda-prl-96}

The second topic, the reversal dynamics in thin ferromagnetic
films finds equally large interest. Both experimental
\cite{raquet-prb-96, lee-prb-99} and theoretical
\cite{raquet-prb-96, wang-prb-99, lyuk-prb-99, ruiz-prb-02}
investigations are performed in order to achieve a better
understanding of the processes during the hysteresis cycle. The
magnetization reversal occurs either by domain wall (DW)
nucleation and motion or by magnetization rotation.\cite{chika}
The DW motion at constant (dc) fields is characterized by three
regions depending on the field strength, that is \emph{creep},
\emph{depinning}, and \emph{sliding motion}. Creep is the
thermally activated motion of DWs, where the average DW velocity
is $v(H) \propto \exp\lbrack -(T_p/T)(H/H_p)^{-\mu} \rbrack$.
\cite{feigel-prl-89, chauve-prb-00, lemerle-prl-97} This behavior
is encountered at small applied fields, $H \ll H_p $, where $H_p$
is the critical depinning threshold and $T_p$ proportional to a
characteristic depinning energy, $U_p=k_B T_p$. At zero
temperature a dynamic phase transition of second order at $H=H_p$
is found. The mean DW velocity, $v$, can be interpreted as order
parameter of the depinning transition, with $v(H) \propto
(H-H_p)^{\beta}$.\cite{roters-pre-99} At $T>0$ the phase
transition is smeared out and the $v(H)$ curve is rounded. Beyond
the depinning region, $H \gg H_p$, sliding motion sets in and the
DW velocity becomes linear with applied field, $v \approx \gamma
H$. Here $\gamma$ is the mobility coefficient.\cite{natt-prl-01,
cowb-apl-99}

In alternating (ac) (magnetic) fields, $H = H_0 \sin(\omega t)$,
additional dynamical effects will arise. The coercive field and
the loop area become dependent on the frequency or in other words
on the field sweep rate,\cite{ruiz-prb-02, lee-prb-99} dynamic
phase transitions and crossovers occur,\cite{natt-prl-01,
lyuk-prb-99, jang-prb-03} the ac susceptibility vs. temperature
shows similar features as spin glass systems \cite{chen-coey-prb}
and a DW velocity hysteresis is found.\cite{glatz-prl-03}
Different models are employed, i.e., numerical solutions of the
coupled differential equations of the DW displacement starting
from Maxwell's equations \cite{wang-prb-99}, using an interface
depinning model for an elastic DW in random
media\cite{glatz-prl-03, natt-prl-01, lyuk-prb-99, stepan-cmat-04}
(sometimes referred to as quenched
Edwards-Wilkinson\cite{edw-wilk} (EW) equation), kinetic
simulations of a DW in the sliding motion regime
\cite{ruiz-prb-02} and calculations based on Fatuzzo's domain
theory\cite{fatuz-pr-62} applied to ultrathin magnetic
layers.\cite{raquet-prb-96}

In this paper we will present model investigations motivated by
recent experiments on the SFM system
$\lbrack$Co$_{80}$Fe$_{20}$(1.4~nm)/
Al$_{2}$O$_{3}$(3~nm)$\rbrack$$_{10}$ being a realization of a
densely packed ensemble of interacting nanoparticles. The complex
magnetic ac susceptibility, $\chi' -i \chi''$, reveals that the
magnetic dynamic behavior can be explained within the concept of
domain wall motion in an impure
ferromagnet.\cite{chen-prl-02,petr-jmmm-04} That means, the
granular system behaves like a thin FM film, only with the
difference, that the atomic moments are to be replaced by
'super-moments' of the individual particles. This concept implies
that the FM nanoparticles remain single-domain whereas the
ensemble shows collective SFM behavior. This idea is evidenced
from the \emph{Cole-Cole plot}, $\chi''$ vs.
$\chi'$.\cite{cole-jcp-41} Hence we will focus on the Cole-Cole
presentation and compare it to that found experimentally.


\section{\label{sec-cole}AC Susceptibility and Cole-Cole Plots}

Magnetic systems exhibiting relaxational phenomena can  be
characterized by the complex ac susceptibility, $\chi(\omega) =
\chi' -i \chi''$. The time dependent complex ac susceptibility is
defined as

\begin{equation}
 M(t)=\tilde\chi(t)\tilde H(t)\,,
\end{equation}
with the {\it complex} external field $\tilde H(t)=-\imath H_0
e^{\imath\omega t}$ [$H(t)=\Re( \tilde H(t))=H_0 \sin(\omega t)$]
and the magnetization $M$. In this paper we study the time
independent term of the Fourier series for $\tilde\chi(t)$,
namely:
\begin{equation}
 \chi\equiv\chi^{\prime}-\imath\chi^{\prime\prime}=\frac{1}{{\cal
 T}}\int\limits_0^{\cal T}dt\,\tilde\chi(t)\,,
\end{equation}
with ${\cal T}=2\pi/\omega=1/f$.

This defines $\chi^{\prime}$ and $\chi^{\prime\prime}$ as follows
\begin{eqnarray}
\label{eqn-acsus1} \chi'(\omega) & = & \frac{1}{H_0{\cal T}}
\int\limits_0^{\cal T}dt\,M(t)\sin(\omega t)
\\\label{eqn-acsus2} \chi''(\omega) & = & -\frac{1}{H_0{\cal T}}
\int\limits_0^{\cal T}dt\, M(t) \cos(\omega t)\,.
\end{eqnarray}
Or equivalently - if we define $\tilde\chi(t)=\frac{dM(t)}{d\tilde
H(t)}=\dot M(t)\left(\frac{d\tilde H}{dt}\right)^{-1}$:
\begin{eqnarray}
\label{eqn-acsus1v} \chi'(\omega) & = & \frac{1}{2\pi H_0}
\int\limits_0^{\cal T}dt\,\dot M(t)\cos(\omega t)
\\\label{eqn-acsus2v} \chi''(\omega) & = & \frac{1}{2\pi H_0}
\int\limits_0^{\cal T}dt\, \dot M(t) \sin(\omega t)\,.
\end{eqnarray}

One way of presenting the data is the Cole-Cole or Argand
representation. The imaginary part is plotted against the real
part of the susceptibility, $\chi''$ vs.
$\chi'$.\cite{cole-jcp-41,jonscher} It can serve as a fingerprint
to distinguish different magnetic systems by their dynamic
response. E.g. a monodisperse ensemble of non-interacting SPM
particles has exactly one relaxation time, $\tau = \tau_0
\exp(KV/k_B T)$, \cite{neel,brown} and will display a semicircle
with the center on the $\chi'$-axis. Here $K$ is an effective
anisotropy constant, $V$ the volume of the particle and $\tau_0$
corresponds to the microscopic spin-flip time which is of order of
$10^{-10}$s. The Cole-Cole plot can easily be derived from an
analytic expression for the ac susceptibility given in
Ref.~\onlinecite{ander-prb-97} for a monodisperse SPM ensemble in
zero-field with a random distribution of anisotropy axis
directions:

\begin{eqnarray}
\chi'(\omega) & = & \mu_0 \frac{M_s^2}{3K} \left[
1+\frac{KV}{k_BT}
\frac{1}{1+(\omega \tau)^2} \right] \\
\chi''(\omega) & = & \mu_0\frac{M_s^2}{3} \frac{V}{k_BT}
\frac{\omega \tau}{1+(\omega \tau)^2}
\end{eqnarray}
where $M_s$ is the saturation value of the magnetization. Defining
$\alpha \equiv \mu_0 M_s^2/3K$ and $\sigma \equiv KV/k_B T$ and
eliminating $\omega$ one gets
\begin{eqnarray}
\chi'' = \sqrt{\left(\frac{\alpha \sigma}{2}\right)^2 -
\left(\chi' - \frac{\alpha (2+\sigma)}{2} \right)^2 } \textrm{,}
\end{eqnarray}
which describes a circle with the radius $r= \alpha \sigma/2$ and
center at $(\alpha (2+\sigma)/2$; $0)$ in the Cole-Cole plane.

\begin{figure}
\includegraphics[width=0.8\linewidth]{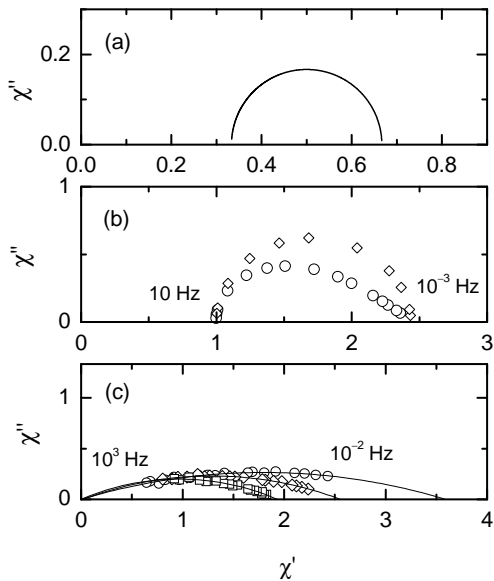}
\caption{\label{fig-colecole} Cole-Cole plots, $\chi''$ vs.
$\chi'$, (a) analytically obtained for a non-interacting
monodisperse ensemble of SPM particles with $\mu_0 M_s^2/3K=1$ and
$KV/k_B T=1$ (see text), (b) numerical result for a polydisperse
ensemble with a log-normal distribution (circles) and a Maxwell
distribution (diamonds) of particle volumes with $\mu_0
M_s^2/3K=1, K/k_B T=1$, $\tau_0=1$, $\Delta V = 0.9$ and $\langle
V\rangle= 1$, and (c) shows experimentally obtained curves on the
SSG system
$\lbrack$CoFe(0.9~nm)/Al$_{2}$O$_{3}$(3~nm)$\rbrack$$_{10}$ at
three different temperatures, $T = 50$, 55 and 60~K
(Ref.~\onlinecite{petr-pt-03}). The particle sizes follow a
Gaussian distribution with $\langle V\rangle$ = 11.5~nm$^3$ and
$\Delta V$ = 0.95~nm$^3$. The frequency range is indicated in the
figure.}
\end{figure}

In Fig.~\ref{fig-colecole}(a) the result is shown for parameters
$\mu_0 M_s^2/3K=1$ and $KV/k_B T=1$. In the case of a particle
size distribution (polydispersivity) and hence a distribution of
relaxation times the Cole-Cole semicircle is expected to become
flattened and/or distorted.\cite{jonscher}
Fig.~\ref{fig-colecole}(b) shows two numerically obtained curves,
where a particle volume distribution from a log-normal
distribution (circles) and a Maxwell distribution\footnote{
$(\sqrt{2} V^2 / \sqrt{\pi} (\Delta V)^3 ) \exp[-V^2/2 (\Delta
V)^2]$.} (diamonds) is assumed using $\mu_0 M_s^2/3K=1, K/k_B
T=1$, $\tau_0=1$, $\langle V\rangle= 1$ and a relatively broad
distribution width $\Delta V = 0.9$. One finds an asymmetric
Cole-Cole plot for the case of a log-normal distribution.
Obviously this is due to the asymmetry of the distribution itself.
By choosing the more symmetric Maxwell distribution the curve
becomes symmetric and only slightly shifted downward. Extremely
high polydispersivity is found in spin glass systems, where the
distribution of relaxation times is expected to become infinitely
broad due to collective behavior.\cite{mydosh}
Fig.~\ref{fig-colecole}(c) shows an experimentally obtained
Cole-Cole plot on the SSG system
$\lbrack$Co$_{80}$Fe$_{20}$(0.9~nm)/
Al$_{2}$O$_{3}$(3~nm)$\rbrack$$_{10}$ at different temperatures,
$T = 50$, 55 and 60~K.\cite{petr-pt-03} Here the particle sizes
follow a relatively narrow Gaussian distribution with $\langle
V\rangle$ = 11.5~nm$^3$ and $\Delta V$ = 0.95~nm$^3$ as evidenced
from a transmission electron microscopy image for a simliar
sample.\cite{sahoo-apl-03}

\section{Models}\label{sec-model}

We study the complex ac susceptibility with two different
approaches, where account is taken of the fact, that $M$ is
controlled by the field-induced sideways motion of one DW. In this
case it follows that $\dot M(t)\propto v(t)$, where $v(t)$ is the
(mean) DW velocity, beeing a function of the external field $H(t)$
and temperature $T$. Both approaches are based on the same
underlying model for a $d$-dimensional elastic DW in a
$D=(d+1)$-dimensional random environment:
\begin{equation}\label{eq.model}
{\cal H}=\int d^dx\,\left\{\frac{\Gamma}{2}\left(\nabla_{\bf
x}Z\right)^2-H(t)Z+V_R({\bf x},Z)\right\}\,,
\end{equation}
where  $Z=Z({\bf x},t)$ is the $d$-dimensional displacement
profile of the DW with internal coordinate ${\bf x}$, $\Gamma$ the
stiffness of the DW, and $V_R$ the (quenched) random potential.
$V_R$ can be written in the following way:
\begin{equation}
 V_R[{\bf x},Z({\bf x},t)]=-\int\limits_0^Z d\tilde Z\,g[{\bf x},\tilde Z({\bf
x},t)]\,,
\end{equation}
where $g[{\bf x},Z({\bf x},t)]$ describes the random force acting
on the DW with $\langle g\rangle=0$ and $\langle g({\bf
x},z)g({\bf x}^{\prime},z^{\prime})\rangle=\delta^d({\bf x}-{\bf
x}^{\prime})\Delta_0(z-z^{\prime})$, with
$\Delta_0(z)=\Delta_0(-z)$ being a random force correlator which
is a monotonically decreasing function decaying over a finite
distance $\ell$.

Since the experimental system is a magnetic film, we restrict
ourselves to the case $D=2$ in the following. The dynamics of the
system follows from the EW equation of motion:
\begin{equation}\label{eq.motion}
 \frac{1}{\gamma}\frac{\partial Z({\bf x},t)}{\partial
 t}=-\frac{\delta {\cal H}}{\delta Z}+\eta({\bf x},t)\,,
\end{equation}
where $\gamma$ is a kinetic coefficient and $\eta({\bf x},t)$ a
thermal noise term. The DW velocity is given by $v({\bf x},t)=\dot
Z({\bf x},t)$. Here we are interested in the mean DW velocity
$v(t)\equiv\langle v({\bf x},t)\rangle_{\bf x}$ and mean
displacement $Z(t)\equiv\langle Z({\bf x},t)\rangle_{\bf x}$, from
which we can calculate the ac susceptibility as described above.
Here $\langle \ldots\rangle_{\bf x}$ denotes the average over the
internal DW coordinate ${\bf x}$.

 {\it (i) adiabatic approach}.--- We use the
expression for the mean DW velocity in the adiabatic driving
regime following from a functional renormalization group (RG)
treatment of (\ref{eq.motion}), given in
Ref.~\onlinecite{natt-prl-01}, which interpolates between the
creep regime and sliding DW motion,

\begin{equation}\label{eq-v}
v(H,T)= \left\{ \begin{array}{ll}
\gamma H F(x,y) & \textrm{for $H \neq 0$,}\\
0 & \textrm{for $H=0$,}
\end{array} \right.
\end{equation}

\noindent where $x = H/H_p$, $y=T_p/T$ and

\begin{eqnarray}
F(x,y)  &=&  \frac{\Theta(1-x)}{1+(yx^{-\mu})^{\beta/\theta}}
\exp\left[yx^{-\mu}(1-x)^{\theta}\right] \\
&&+ \Theta(x-1)\left[
\frac{1}{1+(yx^{-\mu})^{\beta/\theta}}+\left(1-\frac{1}{x}\right)
^{\beta}\right]\,. \nonumber
\end{eqnarray}
Here $\Theta(x)$ is the step function, $T_p\simeq \Gamma
\ell^2L_p^{2-d}$ the typical pinning energy on the Larkin length
scale $L_p$, $H_p$ the zero temperature depinning field, and
$\mu$, $\beta$, and $\theta$ the relevant critical
exponents~\cite{natt-prl-01} which depend on the DW dimension $d$.
A time discretization, $\Delta t$, is used which is chosen to be
much smaller than the period of the driving field, $\Delta t =
10^{-5}{\cal T}$. Then $Z(t)$ is calculated for each time step by
a simple integration of Eq.~\ref{eq-v}, i.e., $\Delta
Z(t_i)=v(H(t_i)) \Delta t$, where $v(H(t_i))$ is calculated for
each time step from Eq.~\ref{eq-v}. Here the values of time $t$,
${\cal T}$, $\omega$, and $f$ are choosen to be dimensionless,
since no quantitative comparison to the experiment is required.
Formally this can be done by introducing an arbitrary time scale
$t_0$ and substituting $t \rightarrow t/t_0$. Analogously this can
be applied to all other parameters and observables, i.e. field
$H_0/H_p \to H_0$, temperature $T/T_p \to T$, velocity $v\to
v/(\gamma H_p)$ and lenght, $L_z/L_0 \to L_z$, where $L_0$ is an
arbitrary lenght scale.

The magnetization for a finite system is defined here as
\begin{equation}\label{magdef}
M(t)=\left(\frac{2Z(t)}{L_z}-1 \right)\,,
\end{equation}
where $L_z$ is the extension of the sample in $Z$-direction and $0
\leq Z \leq L_z$. This implies, that $-1 \leq M \leq +1$. In all
cases the initial condition is $Z(0) = L_z/2$. This approach
includes the temperature as a parameter, but we restrict our
investigations here to small values, $T=0.1$.

{\it (ii) non-adiabatic approach}.--- Since Eq. (\ref{eq-v}) was
obtained for an adiabatically changing field, it can only be used
as an approximation, if the frequency is sufficiently
small\cite{glatz-prl-03} (see also Fig. 2 in this Ref.). In order
to include the pronounced non-adiabatic effects at higher
frequencies (e.g., hysteresis of the velocity), one has to start
with the underlying equation of motion (\ref{eq.motion}) which
yields

\begin{equation}\label{eq-of-mo}
\frac{1}{\gamma}\frac{dZ(\vec{x},t)}{dt}=\Gamma\nabla^2Z(\vec{x},t)+
H(t)+g(\vec{x},Z(\vec{x},t))\,,
\end{equation}
where the thermal noise term is neglected which is justified since
the relaxation times for the DW creep at low temperatures are very
long ($\gg \omega^{-1}$) and we consider only finite (not
exponentially small) frequencies. In
Ref.~\onlinecite{glatz-prl-03} this equation is studied in detail
in the case of an ac driving force in an infinite system and it is
shown that thermal effects are not essential for not too low
frequencies ($\omega>\omega_T\approx\omega_P(T/T_p)^{\nu
z/\theta}$ with the critical exponents $\nu$, $z$, and $\theta$).
Therefore we can restrict ourselves to the zero temperature
equation of motion.

In this approach we investigate both infinite ($L_z \to \infty$)
and finite ($L_z<\infty$) systems. In the second case, the DW will
hit the boundary of the system for low enough frequencies, such
that the magnetization will saturate ($-1\leq M\leq 1$). Therefore
we can derive a critical frequency $\omega_c$ above which the
system will behave as an infinite system. The finite frequency of
the driving force acts as an infrared cutoff for the propagation
of the DW which can move up to a length scale
$L_{\omega}=L_p\left(\Gamma\gamma L_p^{-2}/\omega\right)^{1/z}$.
Equating this scale to $L_z$ gives the following expression for
$\omega_c$ ($f_c$ accordingly)
\begin{equation}
 \omega_c\approx\omega_p(L_p/L_z)^z\,,
\end{equation}
with the typical pinning frequency $\omega_p=\gamma\Gamma/L_p^2$.

For the numerical integration of Eq.~(\ref{eq-of-mo}) it is
discretized in ${\bf x}$-direction(s) into $N^d$ positions with a
lattice constant $\alpha$. Here we also go over to dimensionless
units with an arbitrary time scale $t_0$. These two parameters,
$\alpha$ and $t_0$, are chosen such that
$t_0\gamma\Gamma/\alpha^2=1$ and that the dimensionless random
force $t_0\gamma g$ are set to values in the interval
$\left[-1/2,1/2\right]$ at positions with distance $\ell$. Between
these positions $g$ is interpolated linearly which results in a
Gaussian distribution $\Delta_0(z)$ with variance $\ell$. The
depinning field $H_p$ is not used as input parameter but can be
calculated from Eq.~(\ref{eq-of-mo}) (at $\omega=0$) using a
bisection procedure with constant amplitude.

For our simulations we choose $\ell=0.1$, $N=1000$, for finite
systems $L_z=8.0$, and a time discretization such that $\Delta
t\ll \min(\omega^{-1},0.1)$. The results for $\chi$ are averaged
over $100$ disorder configurations for each frequency.


\section{\label{sec-results}Results and discussion}

\begin{figure}
\includegraphics[width=0.9\linewidth]{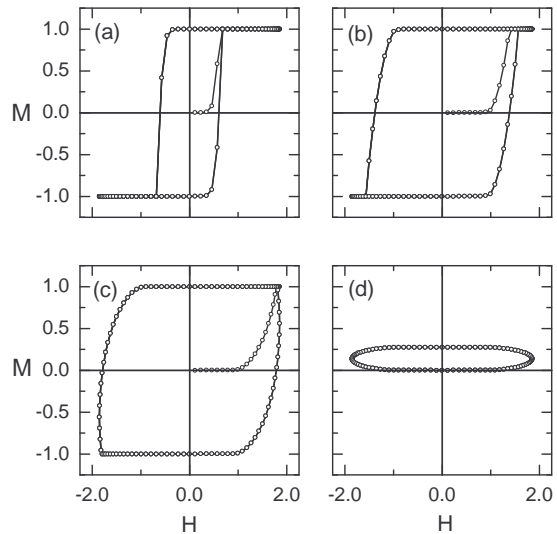}
\caption{\label{fig-hys1} $M$ vs. $H$ curves from simulations of
the adiabatic approach {\it (i)} with $T = 0.1$, $H_0 = 1.85$,
$L_z=8.0$, $\mu=0.24$, $\theta=0.83$, and $\beta=0.66$ at
different frequencies $f=1.6\cdot10^{-7}$ (a), $1.6\cdot10^{-3}$
(b), $7\cdot10^{-3}$ (c) and $8\cdot10^{-2}$ (d). Note that all
quantities are measured in dimensionless units, as described in
the text. Lines are guides to the eye.}
\end{figure}

Fig.~\ref{fig-hys1} shows an example of hysteresis loops from
simulations within approach {\it (i)} with $T = 0.1$, $H_0 = 1.85$
and $L_z=8.0$ at different frequencies $f=1.6\cdot10^{-7}$ (a),
$1.6\cdot10^{-3}$ (b), $7\cdot10^{-3}$ (c) and $8\cdot10^{-2}$
(d). Note that all quantities are measured in dimensionless units,
as mentioned above. For the values of the critical exponents, we
use the results from the RG for $d=2$, i.e., $\mu=0.24$
(Ref.~\onlinecite{lemerle-prl-97}), $\theta=0.83$
(Ref.~\onlinecite{natt-prb-90}), and $\beta=0.66$
(Ref.~\onlinecite{roters-pre-99}) (Note, that the precise values
of these exponents do not have a significant influence on the
behavior under consideration here, especially the qualitative
picture does not change if the values are modified slightly.).
With increasing frequency, the hysteresis loop broadens until it
becomes elliptically shaped above $f=10^{-2}$ loosing also its
inflection symmetry. Similar results are found in
experiments.\cite{chen-prl-02, ruiz-prb-02}

\begin{figure}
\includegraphics[width=0.9\linewidth]{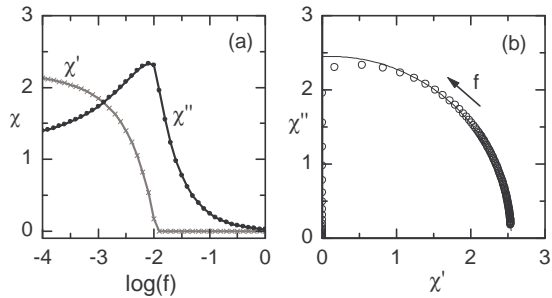}
\caption{\label{fig-sus1} (a) ac susceptibility, $\chi'$ and
$\chi''$ vs. ac frequency, $f$, obtained with model {\it (i)} with
same parameters as in Fig.~\ref{fig-hys1}. (b) Same data, but
plotted in the Cole-Cole presentation, $\chi''$ vs. $\chi'$. The
solid line represents a least square-fit of the low- frequency
data to a circle and the arrow shows in direction of increasing
$f$.}
\end{figure}

The ac susceptibility of such hysteresis cycles can be calculated
from equations (\ref{eqn-acsus1}) and (\ref{eqn-acsus2}). In
Fig.~\ref{fig-sus1} the obtained data is shown for the same set of
parameters as for Fig.~\ref{fig-hys1}. In (a) one finds the real
and imaginary part of the ac susceptibility, $\chi'$ and $\chi''$,
as function of the ac frequency. The real part shows an
order-parameter like behavior with a non-zero value below and a
vanishing value above $f_c\approx 10^{-2}$. Furthermore, the
imaginary part has a peak at $f = 8\cdot10^{-3} \approx f_c$.

\begin{figure}
\includegraphics[width=0.8\linewidth]{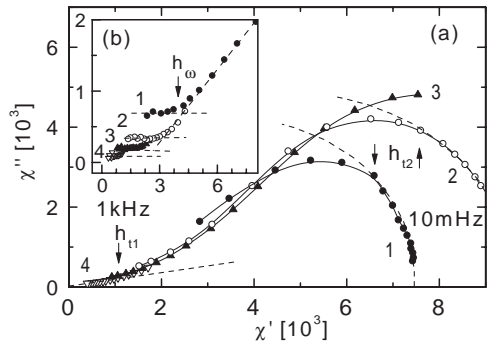}
\caption{\label{fig-chenprl-2} Experimental Cole-Cole plot taken
from Ref.~\onlinecite{chen-prl-02} showing $\chi''$ vs. $\chi'$
obtained on the SFM granular system
[CoFe(1.4nm)/Al$_2$O$_3$(2nm)]$_{10}$. The susceptibility was
mesured at ac amplitudes $\mu_0 H_0 = 50$ (a) and $5 \mu$T (b) at
10~mHz $\leq f \leq$ 1~kHz at $T =$ 380 (1), 350 (2), 320 (3) and
260~K (4). Transition fields are marked by arrows
\cite{chen-prl-02}.}
\end{figure}

In the Cole-Cole plot, Fig.~\ref{fig-sus1} (b), this transition
appears as a sharp change of the slope and curvature. At low
frequencies, $f < f_c$ one observes a quarter-circle centered on
the $\chi'$ axis. It is possible to fit a circle with the center
on the $\chi'$-axis to the low-frequency data (see solid line in
(b)). This corresponds well to the experimental
result~\cite{chen-prl-02,petr-jmmm-04}(Fig.~\ref{fig-chenprl-2})
and suggest the existence of \emph{one} effective relaxation time
in the system. However, for $f>f_c$ only a vertical line can be
observed. This result differs from that found in experiment, where
the high-frequency part is characterized by a positive slope and
positive curvature. This discrepancy needs a closer inspection
here.

By comparison of the susceptibility data to the corresponding
hysteresis loops (Fig.~\ref{fig-hys1}), one sees, that $f_c$ marks
the transition between loops saturating at high fields (low-f) and
those, which do not saturate (high-f). In the second case the
domain wall is always in motion throughout the entire field cycle.
The real part is then zero, whereas the imaginary part has a $1/f$
dependence [Fig.~\ref{fig-sus1} (a)], which follows directly from
our result shown in Ref.~\onlinecite{chen-prl-02}, where the
complex susceptibility in the case of sliding DW motion is given
by $\tilde\chi = \chi_{\infty}[1+1/(i \omega \tau)]$, or more
generally by $\tilde\chi = \chi'_{\infty} + \chi''_{\infty}/(i
\omega \tau)$. For $\chi'_{\infty}=0$ this yields directly the
vertical part in the Cole-Cole plot [Fig.~\ref{fig-sus1} (b)]. In
Ref.~\onlinecite{chen-prl-02} was argued, that the non-linearity
of the $v(H)$ function in the creep regime can be taken into
account by introducing a polydispersivity exponent,$\beta$, in the
above equation, $\tilde\chi = \chi_{\infty}[1+1/(i \omega
\tau)^\beta]$ (compare to a similar relationship formulated for
the conductivity of disordered hopping conductors $\sigma(\omega)
\sim (-i\omega\tau)^{\nu(T)}$, where $0 < \nu < 1$)
\cite{bernasconi-prl-79,ishii-85}. This yields the linear
relationship, $\chi''=\tan(\pi\beta/2)[\chi'-\chi_{\infty}]$.
Note, that for \emph{any} velocity function $v=v(H)$ with
$v(H)=-v(-H)$ and without velocity hysteresis\cite{glatz-prl-03}
it follows, that $\chi'=0$ and $\chi'' \propto
1/f$.\cite{stepan-cmat-04} This can easily be seen from Eqs.
(\ref{eqn-acsus1v}) and (\ref{eqn-acsus2v}) and $\dot M\propto v$.
The consequence is that a monotonically increasing part with
finite slope cannot be found in the Cole-Cole plot by considering
only the adiabatic motion of one DW.

There are two possible ways to improve the model. The first one is
to simulate an ensemble of non-interacting subsystems with
different domain propagations lengths, pinning fields, $H_p$, or
depinning energies, $T_p$. It is possible, that this case would
yield the situation above qualitatively described by the
polydispersivity exponent $\beta<1$. The second is to employ a
more realistic description of the DW by using the above introduced
non-adiabatic approach {\it (ii)}. The latter case was studied
here.

\begin{figure}
\includegraphics[width=0.9\linewidth]{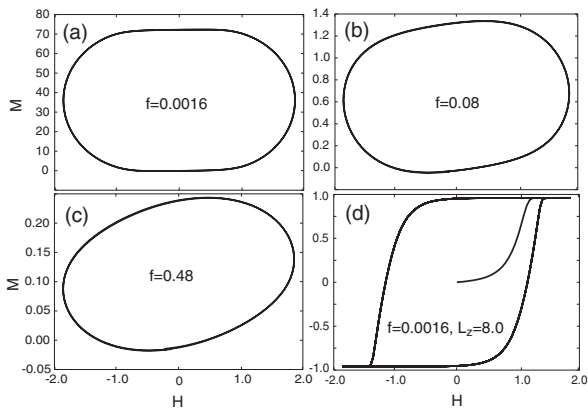}
\caption{\label{fig-hys2} $M$ vs. $H$ curves for the non-adiabatic
approach {\it (ii)} with $H_0=1.85$ at different frequencies,
$f=0.0016$ (a), 0.08 (b), and 0.48 (c) for an infinite system (in
this case one defines $M=Z$). (d) shows the magnetization curve
for $f=0.0016$ but a finite system ($L_z=8.0$) so that the DW
touches the boundaries. }
\end{figure}

In Fig.~\ref{fig-hys2} the results for the magnetization
hysteresis of a DW from Eq.~\ref{eq-of-mo} for $H_0=1.85$ are
presented. The plots (a) to (c) show hysteresis loops at different
frequencies, $f=0.0016$ (a), 0.08 (b), and 0.48 (c) for an
infinite system ($L_z\to\infty$). Here we define $M=Z$. In this
case the DW never touches the sample boundary. At low frequencies
one finds a symmetric loop with respect to the $M$ axis (a)
similar to the result shown above in Fig.~\ref{fig-hys1}(d). This
symmetry is lost upon increasing the frequency, (b) and (c), and
the loop becomes tilted. This tilting is responsible for a
non-vanishing real part of the ac susceptibility and cannot be
observed in approach {\it (i)}. The tilting corresponds to the
appearance of a velocity hysteresis.\cite{glatz-prl-03} That means
there exists no functional relationship between the velocity and
the field any more, as it is the case in the adiabatic regime.

\begin{figure}
\includegraphics[width=0.9\linewidth]{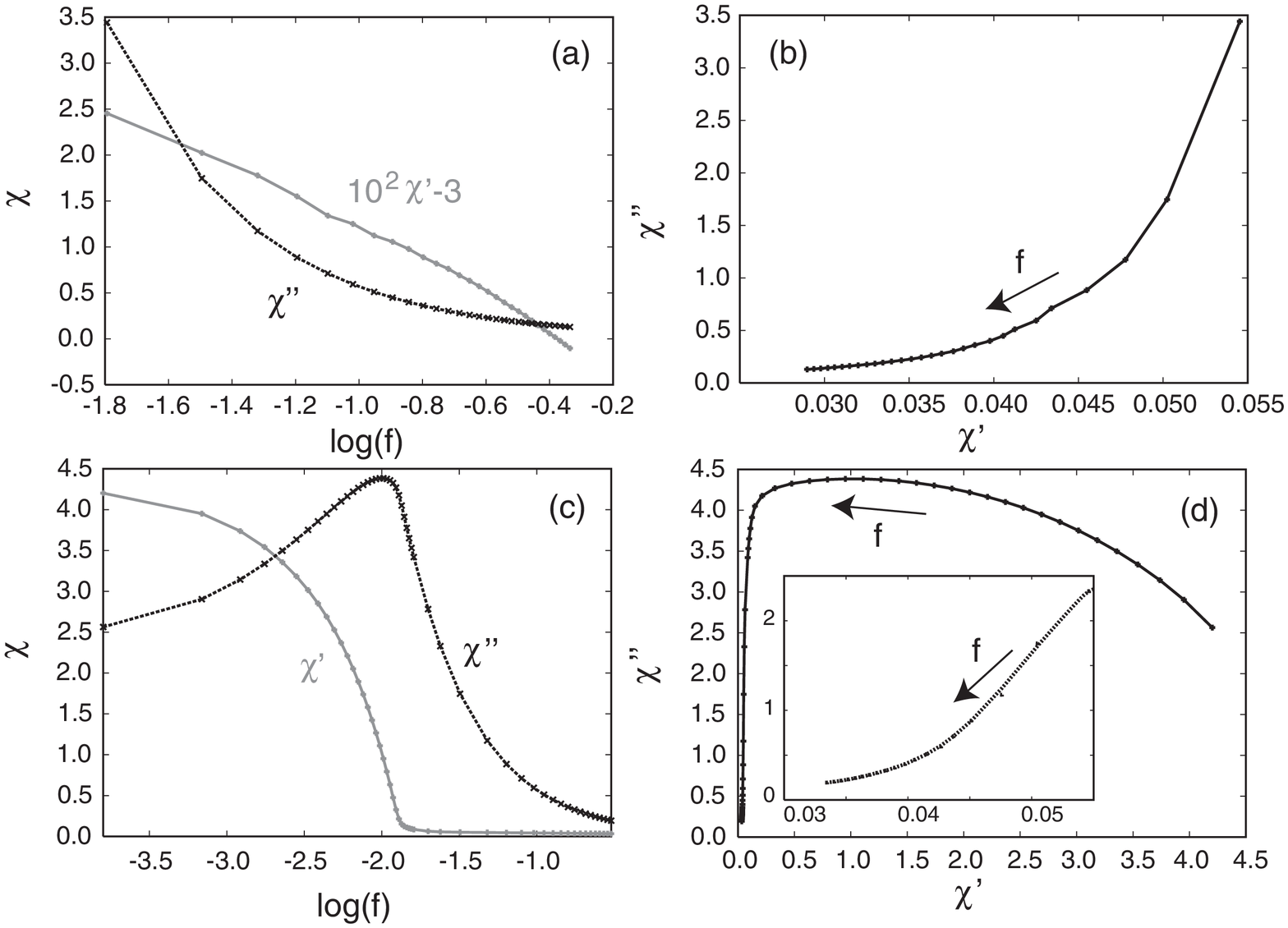}
\caption{\label{fig-sus2} Real and Imaginary part of the ac
susceptibility vs. frequency, calculated within approach {\it
(ii)} for infinite systems (a) [the real part is shifted and
scaled] and the corresponding Cole-Cole plot (b). In (c)
$\chi^{\prime}$ and $\chi^{\prime\prime}$ are plotted for the
finite system ($L_z=8.0$) and Cole-Cole plot (d). The inset in (d)
shows the high frequency behavior in more detail. The arrows show
in direction of increasing frequencies. All simulations were
performed with $H_0=1.85$. }
\end{figure}

The resulting susceptibilities are plotted in Fig.~\ref{fig-sus2}.
In (a) and (b) the real and imaginary part vs. $\log(f)$ and the
corresponding Cole-Cole plot, respectively, are shown for an
infinite system, when the DW never touches the boundary. In (c)
and (d) the same plots are shown for a finite system ($L_z=8.0$).
While the low-frequency parts resemble those from approach {\it
(i)}, the high-frequency part shows a completely different
behavior. For $\chi' \rightarrow 0$ we find in the Cole-Cole plot
[inset in Fig.~\ref{fig-sus2} (d)] a curve with positive curvature
similar as in the experiment (Fig.~\ref{fig-chenprl-2}). One can
expect that $\chi$ goes to $0$ with $\omega\to\infty$, since the
velocity hysteresis disappears for $\omega\to\infty$. Obviously
the more realistic second model is capable to describe the
experimentally found behavior. At this point we want to emphasize
that the adiabatic approach only works for low frequencies, where
non-adiabatic effects can be neglected. Furthermore it only works
at finite temperatures. On the other hand the non-adiabatic
approach can explain the main experimental features even if we use
the zero temperature equation of motion, since the smearing
effects of the depinning transition due to finite frequencies
dominate the thermal creep effects at low temperatures.

However, two drawbacks still exist. One, the Cole-Cole plot from
the simulation shows a rather steep and narrow increasing part
compared to the experiment. Second, we cannot retrieve the
experimentally observed saturating part for the highest
experimental frequencies, where the imaginary part becomes
constant (see Fig.~\ref{fig-chenprl-2}, inset). This case was
attributed~\cite{chen-prl-02} to the reversible relaxation
response of the DW for high frequencies and small excitations
fields.\cite{stepan-cmat-04,natt-prb-90} It would be interesting
to study this case with a suitably modified model which includes
multiple and interacting DWs.

\section{\label{sec-conclusion}Conclusion}

In order to get a better understanding of the magnetic behavior
found in the superferromagnetic granular multilayer
$\lbrack$Co$_{80}$Fe$_{20}$(1.4nm)/Al$_{2}$O$_{3}$(3nm)$\rbrack$$_{10}$
we employed two types of simulations of a domain wall in random
media driven by an external magnetic field. Using the first
approach with the mean velocity of a domain wall in the adiabatic
limit, one can explain the monodisperse dynamic response evidenced
by a partial semicircle centered around the $\chi'$ axis. However,
it fails to describe the increasing part with positive curvature
for higher frequencies in the Cole-Cole plot. This behavior can be
found by taking the full equation of motion into account, where an
elastic interface is driven in general {\it non-adiabatically} in
a random medium. Hence a model of an impure ferromagnet is capable
to describe the main features of the experimental results. We find
that the appearance of a velocity hysteresis is a crucial element
in the dynamic response of the superferromagnet. We show that a
Cole-Cole plot may be used to classify magnetic systems by their
dynamic response. E.g. the above mentioned granular
superferromagnet can unambiguously be distinguished from a
superparamagnet and a superspin glass system.


\begin{acknowledgments}
We thank Ch. Binek and X. Chen for helpful discussions. This work
has been supported by the Alexander-von-Humboldt Foundation (O.P.)
and the Deutscher Akademischer Austauschdienst (A.G.).
\end{acknowledgments}



\bibliography{Petracic_Glatz_v4}
\clearpage


\end{document}